\documentclass[reprint,aps]{revtex4-1}
\usepackage{bbm}
\usepackage{amsmath}
\usepackage{amssymb}

\usepackage{epsfig}
\usepackage{graphicx}
\usepackage{color}
\usepackage{mathrsfs}
\usepackage{amsfonts}
\usepackage{multirow}
\usepackage{color}
\usepackage{subfigure}

\begin{document}

\title{ Numerical variational simulations of quantum phase transitions in the sub-Ohmic spin-boson model with multiple polaron ansatz}

\author{Yulong Shen and  Nengji Zhou}
\email[Electronic address: ]{zhounengji@hznu.edu.cn}
\date{\today}
\affiliation{Department of Physics, Hangzhou Normal University, Hangzhou 311121, China
}

\begin{abstract}
With extensive variational simulations, dissipative quantum phase transitions in the sub-Ohmic spin-boson model are numerically studied in a dense limit of environmental modes. By employing a generalized trial wave function composed of coherent-state expansions, transition points and critical exponents  are accurately determined for various spectral exponents, demonstrating excellent agreement with those obtained by other sophisticated numerical techniques. Besides, the quantum-to-classical correspondence is fully confirmed over the entire sub-Ohmic range, compared with theoretical predictions of the long-range Ising model. Mean-field and non-mean-field critical behaviors are found in the deep and shallow sub-Ohmic regimes, respectively, and distinct physical mechanisms of them are uncovered.
\end{abstract}

\maketitle

\section{Introduction}
\begin{figure*}[htbp]
  \centering
   \includegraphics[scale=0.4]{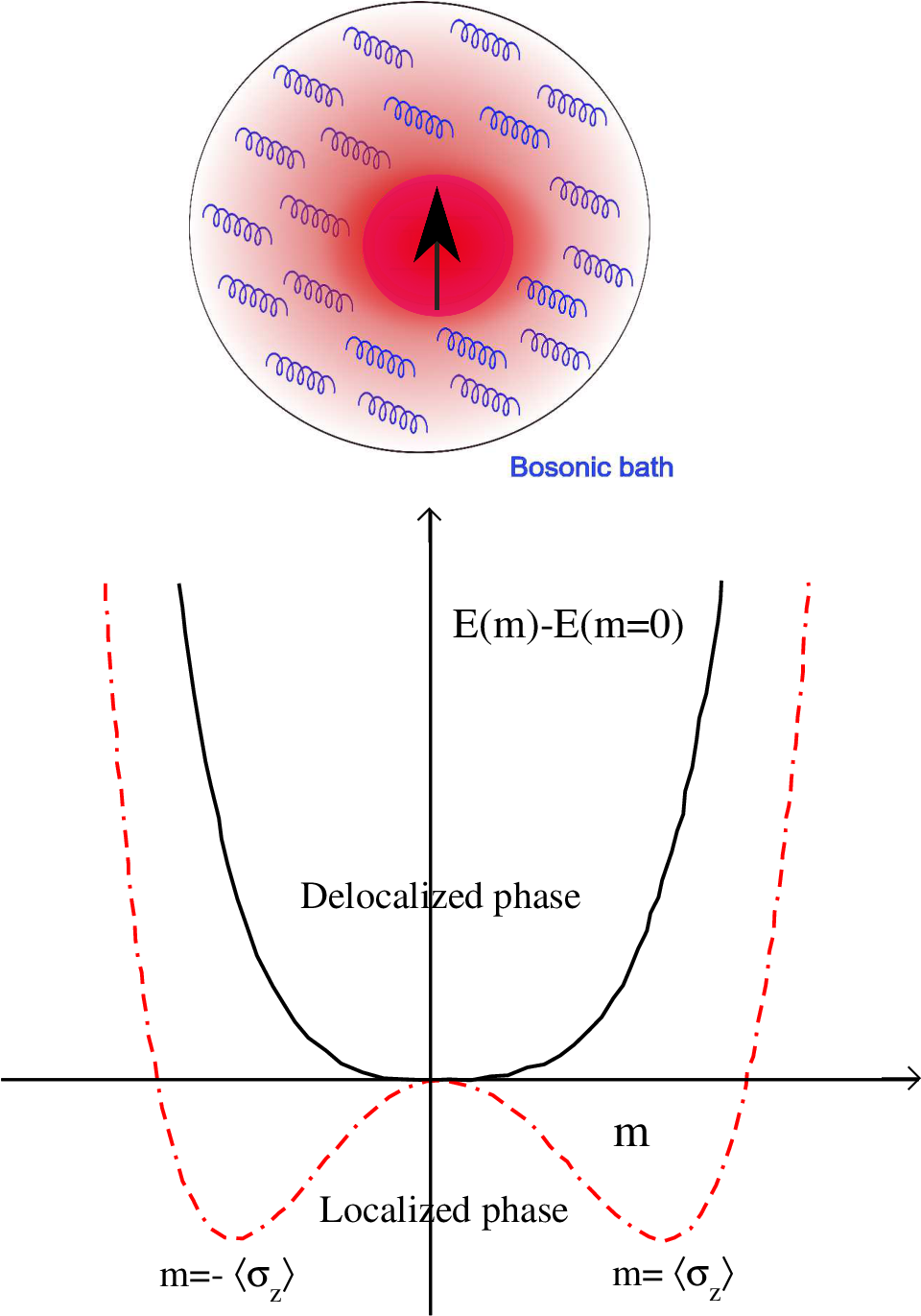}
   \caption{ Schematic of SBM where a quantum spin is immersed in a bosonic bath (environment) represented by a collection of harmonic oscillators. The sketch of the quantum phase transition for the unbiased SBM is shown at the bottom, which separates a non-degenerate delocalized phase from a doubly degenerate localized one. The system energy difference $E(m)-E(m=0)$ $vs.$ the magnetization $m$ is plotted for these two phases, and the ground-state values of the magnetizations $m=\pm \langle\sigma_z\rangle$ and $0$ are obtained as the local minima of the energy landscape, respectively. }
   \label{f1}
\end{figure*}

In recent years, there has been significant theoretical and experimental interest in quantum phase transitions occurring at absolute zero temperature, which are driven by quantum fluctuations that can be tuned by the external magnetic field or the coupling constant \cite{voj03,hur10,sac11}. At low temperatures, they can still be detected if the system is dominated by quantum fluctuations \cite{son97,pai05,wer10}. Most of the works on quantum phase transitions have focused on closed quantum systems. However, it is also important to consider the impact of inherent system-environment coupling on physical features such as decoherence, dissipation, and entanglement \cite{wei07}. As a result, quantum critical phenomena in dissipation systems have recently received considerable attention \cite{ort10,ban14,mai18,ros21}. The spin-boson model (SBM), as schematically depicted in the top panel of Fig.~\ref{f1}, is a prominent example featuring a two-level system (quantum spin) interacting with harmonic oscillators \cite{leg87,hur08,bre16}. Despite its apparent simplicity, the SBM captures the physics of a diverse range of systems, from defects in solids to quantum thermodynamics, physical chemistry, and biological systems \cite{lew88,gol92,cha95,eng07,por08,col09,ota11,uzd15}. Recent studies have shown that the SBM and its variants exhibit rich ground-state and dynamic phase transitions with the increasing system-environment coupling \cite{guo12,nal13,zho14,wan16,zho18,wan20,per23}.

Gapless bosonic baths characterized by a power-law spectral density $J(\omega) \propto  \omega^s$ can be divided into three distinct regimes depending on the value of the spectral exponent $s$: sub-Ohmic ($s < 1$), Ohmic ($s =1$), and super-Ohmic ($s > 1$) \cite{leg87}. Due to the competition between the tunneling and environment-induced dissipation, the SBM undergoes a quantum phase transition which separates a non-degenerate delocalized phase from a doubly degenerate localized phase \cite{keh96,hur08}, as illustrated in the bottom panel of Fig.~\ref{f1}. The phase transition is of second order in the sub-Ohmic regime and takes on the Kosterlitz-Thouless type in the Ohmic regime. In the super-Ohmic regime, however, there is no phase transition. In addition, the quantum-classical mapping predicts that the sub-Ohmic SBM should be equivalent to a classical Ising spin chain with long-range interactions, and mean-field critical exponents are expected in the deep sub-Ohmic regime with $0<s\le 0.5$ \cite{leg87,bul03,ort10,guo12}.

Although an exact analytical solution is currently lacking, various powerful numerical approaches exist to investigate ground-state properties of the SBM. Among them, the most famous one is the Numerical Renormalization Group (NRG) developed by Kenneth Wilson in the $1970$s \cite{wil71}. Failure of the quantum-to-classical correspondence was reported in NRG works \cite{bul03,voj05}, since the results of critical exponents did not align with the mean-field values.  Conversely, the mapping was confirmed by other numerical approaches, such as Quantum Monte Carlo (QMC) \cite{win09}, Density-Matrix Renormalization Group (DMRG) \cite{won08}, Exact Diagonalization (ED) \cite{alv09}, and Variational Matrix Product States (VMPS) \cite{guo12}. The controversy was subsequently addressed for the realization that the earlier NRG results were incorrect due to bosonic state space truncation \cite{ton12}. Besides, the ground-state phase transition can also be investigated with quantum dissipation dynamic approaches \cite{zho08,yan16,dua17,wan19}.

In the shallow sub-Ohmic regime with $0.5 < s < 1$, $s$-dependent values of critical exponents were obtained going beyond the mean-field theory \cite{voj05,win09,blu17,bru17}. It suggests that further examination on the validity of the quantum-to-classical correspondence is required. However, few research has been conducted except for that carried out by VMPS \cite{guo12}. Besides, numerical results in such a regime show significant variations in the critical couplings. For instance, the NRG one is greater than others by nearly $10$ percent at $s = 0.9$ \cite{alv09,win09,zha10}. Very recently, quantum simulations on the SBM have been realized in experiments of superconducting quantum circuits and trapped atomic ion crystals \cite{por08,lep18,mag18,lem18}, offering the chance of experimental studies on quantum phase transitions. However, current understanding of quantum criticality in the SBM remains limited, particularly in the context of the shallow sub-Ohmic regime.

The variational technique is another popular method of obtaining approximate energies and wave functions in quantum mechanical systems \cite{he99,sor05}, where the form of the trial wave function is crucial. The original variational work on the SBM utilized the polaronic unitary transformation proposed by Silbey and Harris \cite{sil84}. This ansatz was later adapted into an asymmetric form by Chin et al$.$ to investigate quantum phase transitions in the sub-Ohmic SBM \cite{chi11}. Mean-field critical properties were found in the deep sub-Ohmic regime, consistent with those obtained from other numerical methods. However, the investigation of cases within the shallow sub-Ohmic regime was unexplored. With the ansatz modified by superposing more than one nonorthogonal coherent states, the mean-field value of the magnetization critical exponent $\beta=0.50(1)$ was confirmed recently not only for $0<s\le0.5$, but also for $0.5<s<1$ \cite{he18}. But, unexpectedly, the $\beta$ values in the latter differ greatly from VMPS ones which are much less than $0.5$ \cite{guo12,fre13}, thus casting doubt on the ability of the variational method to treat quantum phase transitions across the entire sub-Ohmic regime.

In this article, extensive variational calculations with thousands of parameters are carried out to study quantum phase transitions in both deep and shallow sub-Ohmic SBM based on the generalized trial wave function composed of coherent-state expansions, which has been proved to be valid in tackling ground-state and dynamic properties \cite{zho14,zho15,zho15b,zho16,zho18,zho21,zho22}. In the continuum limit, transition points and critical exponents are accurately determined, and the variational results are examined in detail, in comparison with those obtained from other numerical approaches. Moreover, the quantum-to-classical correspondence is confirmed in the whole sub-Ohmic regime, and the nature of the mean-field and non-mean-field behaviors as well as the finite-size effect are also analyzed. The rest of the paper is organized as follows. In Sec.~\ref{sec:mod}, the model, variational approach and scaling analysis are described, and in Sec.~\ref{sec:num}, the numerical results are presented. Finally, Sec.~\ref{sec:con} includes the conclusion.

\section{Methodology}\label{sec:mod}
\subsection{Model and Method}
Numerical studies of quantum phase transitions are performed in the SBM whose Hamiltonian can be described by
\begin{equation}
\label{Ohami_single}
\hat{H} =  \frac{\varepsilon}{2}\sigma_z-\frac{\Delta}{2}\sigma_x + \sum_{k} \omega_k b_{k}^\dag b_{k}  +  \frac{\sigma_z}{2}\sum_k \lambda_k(b^\dag_{k}+b_{k}),
\end{equation}
where $\varepsilon$ ($\Delta$) denotes the energy bias (tunneling amplitude), $b^\dag_k$ ($b_k$) represents the bosonic creation (annihilation) operator of the $k$-th bath mode with the frequency $\omega_k$, $\sigma_x$ and $\sigma_z$ are Pauli spin-$1/2$ operators, and $\lambda_k$ signifies the coupling amplitude between the system and bath. To simplify notation, we fix the Planck's constant $\hbar =1$, so model parameters $\Delta, \varepsilon, \lambda_k$, and $\omega_k$ are dimensionless. With the coarse-grained treatment based on the Wilson energy mesh, the values of $\lambda_k$ and  $\omega_k$ in Eq.~(\ref{Ohami_single}) can be calculated from the continuous spectral density function $J(\omega)=2\alpha\omega_c^{1-s}\omega^s =\sum_k\lambda_k^2\delta(\omega-\omega_k)$ \cite{bul05, voj05,zha10,zho14, blu17}, where $\alpha$ denotes the dimensionless coupling strength, and $\omega_c=1$ represents the high-frequency cutoff. In order to obtain an accurate description of the ground state for SBM in a high dense spectrum, the factor $\Lambda=1.05$ is set in the logarithmic discretization procedure. It is much closer to the continuum limit $\Lambda \rightarrow 1$, compared to those in previous numerical works \cite{bul03,hur08,chi11,fre13,ber14b}. Furthermore, the convergency test of the logarithmic discretization factor is also performed, and the results demonstrate that a value of $\Lambda=1.05$ is sufficient to obtain reliable outcomes \cite{zho22}.

The multiple polaron ansatz \cite{zho14}, which is also known as Davydov multi-D$_1$ ansatz, is used in the studies with the numerical variational method (NVM),
\begin{eqnarray}
\label{vmwave1}
|\Psi_g \rangle & = & | \uparrow \rangle \sum_{n=1}^{N} A_n \exp\left[ \sum_{k=1}^{M}\left(f_{n,k}b_k^{\dag} - \mbox{H}.\mbox{c}.\right)\right] |0\rangle_{\textrm{b}} \nonumber \\
              & + & |\downarrow \rangle \sum_{n=1}^{N} B_n \exp\left[ \sum_{k=1}^{M}\left(g_{n,k}b_k^{\dag} - \mbox{H}.\mbox{c}.\right)\right] |0\rangle_{\textrm{b}},
\end{eqnarray}
where H$.$c$.\!$ denotes Hermitian conjugate, $| \uparrow \rangle$ ($| \downarrow \rangle$) stands for the spin up (down) state, and $|0\rangle_{\rm b}$ is the vacuum state of the bosonic bath. The variational parameters $f_{n,k}$ and $g_{n,k}$ represent displacements of the coherent states correlated to spin configurations, and $A_n$ and $B_n$ are weights of coherent states. The subscripts $n$ and $k$ correspond to the ranks of the coherent superposition state and effective bath mode, respectively.

The ground state  $|\Psi_g\rangle$ is obtained by minimizing the energy expressed as $E=\mathcal{H}/\mathcal{N}$, using the Hamiltonian expectation $\mathcal{H}=\langle \Psi_g|\hat{H}|\Psi_g\rangle$ and the norm of the wave function $\mathcal{N}=\langle \Psi_g |\Psi_g\rangle$,
\begin{eqnarray}
&& \mathcal{H} =          \nonumber \\
&& \sum_{m,n}A_mB_n\Gamma_{m,n}(-\Delta) + \frac{\varepsilon}{2}\sum_{m,n}\left(A_mA_n F_{m,n} - B_mB_n G_{m,n}\right)  \nonumber \\
+ && \sum_{m,n}A_mA_n F_{m,n}\sum_{k}\left[\omega_{k}f_{m,k}f_{n,k} + \frac{\lambda_{k}}{2}(f_{m,k}+f_{n,k})\right]   \\
+ && \sum_{m,n}B_mB_n G_{m,n}\sum_{k}\left[\omega_{k}g_{m,k}g_{n,k} - \frac{\lambda_{k}}{2}(g_{m,k}+g_{n,k})\right], \nonumber
\label{vmH}
\end{eqnarray}
and
\begin{equation}
\mathcal{N} = \sum_{m,n}\left( A_mA_nF_{m,n} + B_mB_nG_{m,n}\right),
\label{vmD}
\end{equation}
where $F_{m,n},G_{m,n}$, and $\Gamma_{m,n}$ are Debye-Waller factors defined as
\begin{eqnarray}
F_{m,n} & = & \exp\left[-\frac{1}{2}\sum_{k}(f_{m,k}-f_{n,k})^2\right], \nonumber \\
G_{m,n} & = & \exp\left[-\frac{1}{2}\sum_{k}(g_{m,k}-g_{n,k})^2\right],  \\
\Gamma_{m,n} & = & \exp\left[-\frac{1}{2}\sum_{k}(f_{m,k}-g_{n,k})^2\right]. \nonumber
\label{vmfactor}
\end{eqnarray}
A set of self-consistency equations are then deduced by the Lagrange multiplier method,
\begin{equation}
\frac{\partial \mathcal{H}}{\partial x_{i}} - E\frac{\partial\mathcal{N}}{\partial x_{i}} = 0,
\label{vmit}
\end{equation}
with respect to the variational parameter $x_i(i=1,2,3,\cdots, 2NM+2N)$. Finally, the iterative equations are derived,
{\small
\begin{eqnarray}
A_n^{*} & = &  \frac{\sum_{m}B_m\Gamma_{n,m}(-\Delta)+2\sum_m^{m\neq n}A_mF_{m,n}(aa_{m,n}-E)}{2(E-aa_{n,n})}, \nonumber \\
B_n^{*} & = &  \frac{\sum_{m}A_m\Gamma_{m,n}(-\Delta)+2\sum_m^{m\neq n}B_mG_{m,n}(bb_{m,n}-E)}{2(E-bb_{n,n})},  \nonumber \\
f_{m,k}^{*} & = & \frac{2\sum_n^{n\neq m} A_nF_{m,n}(\omega_kf_{n,k}+\lambda_k/2+aa_{m,n}f_{n,k}-Ef_{n,k})}{2A_m(E-\omega_k-aa_{m,m})}  \nonumber \\
        & + &  \frac{\sum_nB_n\Gamma_{m,n}g_{n,k}(-\Delta) + A_m\lambda_k}{2A_m(E-\omega_k-aa_{m,m})},   \\
g_{m,k}^{*} & = & \frac{2\sum_n^{n\neq m} B_nG_{m,n}(\omega_kg_{n,k}-\lambda_k/2+bb_{m,n}g_{n,k}-Eg_{n,k})}{2B_m(E-\omega_k-bb_{m,m})}  \nonumber \\
        & + &  \frac{\sum_nA_n\Gamma_{n,m}f_{n,k}(-\Delta) - B_m\lambda_k}{2B_m(E-\omega_k-bb_{m,m})}, \nonumber
\label{vmit2}
\end{eqnarray}
}
where $aa_{m,n}$ and $bb_{m,n}$ denote
\begin{eqnarray}
aa_{m,n} & = & \sum_k\left[\omega_kf_{m,k}f_{n,k} + \frac{\lambda_k}{2}(f_{m,k}+f_{n,k})\right] + \frac{\varepsilon}{2},   \nonumber \\
bb_{m,n} & = & \sum_k\left[\omega_kg_{m,k}g_{n,k} - \frac{\lambda_k}{2}(g_{m,k}+g_{n,k})\right] - \frac{\varepsilon}{2},
\label{vmfactor2}
\end{eqnarray}
respectively. Using the relaxation iteration technique, one updates the variation parameter by $x_i'=x_i+t*(x_i^{*}-x_i)$ where $x_i^{*}$ is defined in Eq.~(\ref{vmit2}), and $t=0.1$ is the relaxation factor. The termination criterion of the iteration procedure is set to be max$\{x_i^{*} -x_i\} < 10^{-13}$ over all variational parameters. More than one hundred random initial states are taken to reduce statistical noise for each set of model parameters ($\alpha, \Delta, \Lambda, \varepsilon$). Furthermore, simulated annealing algorithm is also employed to escape from metastable states.

With the ground-state wavefunction $|\Psi_g\rangle$ at hand, the spin magnetization $\langle\sigma_{z}\rangle$ and spin coherence $\langle\sigma_{x}\rangle$ can be measured by
\begin{eqnarray}
\langle\sigma_z\rangle & = &\frac{\sum_{m,n}\left(A_mA_nF_{m,n} - B_mB_nG_{m,n}\right)}{\sum_{m,n}\left(A_mA_nF_{m,n} + B_mB_nG_{m,n}\right)}.  \nonumber \\
\langle\sigma_x\rangle & = &\frac{\sum_{m,n}2A_mB_n\Gamma_{m,n}}{\sum_{m,n}A_mA_nF_{m,n}+B_mB_nG_{m,n}}.
\label{vm_mag}
\end{eqnarray}
The local susceptibility $\chi$ is also investigated for a nonvanishing bias $\varepsilon$,
\begin{equation}
\chi (\alpha) = \left.\frac{\partial \langle\sigma_z\rangle}{\partial \varepsilon} =\frac{\delta \langle\sigma_z\rangle}{\delta \varepsilon}\right|_{\varepsilon = 0},
\label{vm_sup}
\end{equation}
where the variation of the spin magnetization $\delta \langle\sigma_z\rangle$ is calculated under $\delta \varepsilon=10^{-5}\omega_c$.

Besides spin-related observations, a many-body quantum tomography technique focused on the bosonic bath is also introduced, which allows direct characterization of the ground-state wave function. It can be accessed via standard Wigner tomography {\cite{lvo09}}, which is a method used to reconstruct the Wigner function of the quantum system. It allows for the characterization of the complete quantum state. The basic principle is to perform a series of measurements that directly probe the position and momentum values of the quantum system. By repeating this measurement and reconstruction process for different measurement settings, one can build up a comprehensive picture of the quantum state's Wigner function in the phase space. This provides insights into the system's quantum coherence, entanglement, and non-classical behavior. In this work, the Wigner function $W^k(X,P)$ can be calculated after all the degrees are traced out except the single bath mode with the quantum number $k$. Thus, the Wigner function $W^k(\beta)$ is introduced in the well-known definition \cite{ber14b},
\begin{eqnarray}
W^k(\beta) & = & \frac{1}{\mathcal{N}}\int \frac{d^2\eta}{\pi^2}e^{-(\eta\beta^* - \eta^*\beta)} \mbox{Tr}\left[\hat{\rho}e^{\eta b_k^{\dag} - \eta^*b_k}\right] \nonumber \\
&=& \frac{1}{\mathcal{N}}\int \frac{d^2\eta}{\pi^2}e^{-(\eta\beta^* - \eta^*\beta)}\langle\Psi_g|e^{\eta b_k^{\dag} - \eta^*b_k}|\Psi_g\rangle \nonumber \\
 & = & \frac{1}{\mathcal{N}}\int\frac{d^{2}\eta}{\pi^{2}}\exp\left(\eta^*\beta-\eta\beta^{*}-\eta\eta^*/2\right)\\
 & \times & \sum_{m,n}\left[A_{m}A_{n}F_{mn}\exp\left(\eta f_{m,l}-\eta^*f_{n,l}\right)\right. \nonumber \\
 & + & \left. B_{m}B_{n}G_{mn}\exp\left(\eta g_{m,l}-\eta^*g_{n,l}\right)\right], \nonumber
\label{vm_wigner}
\end{eqnarray}
where $\beta=X+iY$ is a complex number. Taking the Gaussian integral over the variable $\eta$ in the complex space as well as the variable $Y$, one can obtain
\begin{eqnarray}
W^k(X) & = & \frac{\sqrt{2}}{\sqrt{\pi}\mathcal{N}} \sum_{m,n}A_mA_n \exp\left[-\frac{1}{2}\sum_{l\neq k}(f_{m,l}-f_{n,l})^2\right] \nonumber \\
&\times&\exp\left[-2\left(X-(f_{m,k}+f_{n,k})/2\right)^2\right]  \\
&+& \frac{\sqrt{2}}{\sqrt{\pi}\mathcal{N}} \sum_{m,n}B_mB_n \exp\left[-\frac{1}{2}\sum_{l\neq k}(g_{m,l}-g_{n,l})^2\right] \nonumber \\
&\times&\exp\left[-2\left(X-(g_{m,k}+g_{n,k})/2\right)^2\right]. \nonumber
\label{vm_wigner2}
\end{eqnarray}
In fact, the function $W^k (X)$ represents the probability distribution of the $k$-th bath mode in the position space, where the position operator is $\hat{x}=(b_k+b_k^\dag )/\sqrt{2}$.

\subsection{Implementation of the algorithm}
In the following, the algorithm of the numerical variational method in this paper is implemented with the architecture:

Step $1$: Initialize variational parameters randomly. Specifically, variational parameters $(A_n,B_n)$ are uniformly distributed within
an interval $[-1, 1]$, and displacement coefficients $(f_{m,k}, g_{m,k})$ of the initial states obey $f_{m,k}=-g_{m,k} \sim \lambda_k/2(\omega_k+0.01\omega_r)$
with a uniformly distributed random frequency $\omega_r$.

Step $2$:  Update variational parameters with the relaxation iteration technique until the preliminary condition max$\{x_i^{*} -x_i\} < 10^{-8}$ over all variational parameters is reached.
The trail wavefunction and its energy are then recorded.

Step $3$:  Repeat the steps $1$ and $2$ for more than $100$ times, and subsequently choose the trail wavefunction with the lowest energy as the candidate for the ground state.

Step $4$:  Carry on the iterative procedure in the candidate until the target precision $10^{-13}$ is reached. If necessary, the simulated annealing algorithm is also used where the relaxation factor gradually decreases to $t=0.001$ in order to improve the energy minimization procedure.

Step $5$:  Measure the ground-state properties with the spin magnetization $\langle\sigma_{z}\rangle$, the spin coherence $\langle\sigma_{x}\rangle$, the local susceptibility $\chi$, and the Wigner function $W^k(X)$ in Eqs.~(\ref{vm_mag})-(\ref{vm_wigner}).

The FORTRAN source codes that we developed, which allow for the treatment of quantum phase transitions in the sub-Ohmic SBM, are provided as the supplementary material. Moreover, the Central Processing Unit (CPU) time and cost of the memory for each set of model parameters are also estimated, which are about $2.5$ months on a single processor and $2$ megabytes of the memory. The type of CPU is ``Intel Xeon Gold $6248$R'' with the frequency $3.00$GHz, and we have $176$ logical processors in total on a Linux-based computing cluster.

\subsection{Scaling analysis}
Since the localized-delocalized phase transition in the sub-Ohmic SBM is of second order, the order parameter $\langle\sigma_z\rangle$ and local susceptibility $\chi$ should obey
\begin{equation}
\label{scaling law}
\langle\sigma_z\rangle \sim \tau^\beta,\qquad \langle\sigma_z\rangle \sim |\varepsilon|^{1/\delta}, \qquad \chi\sim \tau^{-\gamma},
\end{equation}
where the reduced coupling $\tau=|\alpha-\alpha_c|/\alpha_c$ denotes the distance from the critical coupling, and $\beta, \gamma$, and $\delta$ are critical exponents. Analogous to a classical phase transition, the correlation length in imaginary time diverges as a function of the distance, $\xi \sim \tau^{-\nu}$, where $\nu$ is the correlation-length exponent. Thus, the transition point can be determined by $\delta\alpha_c(L)=\alpha_c(L)-\alpha_{c,\infty} \sim L^{-1/\nu}$ with the finite size $L$ of the system. Note that the correlation length is given by an inverse energy scale $\xi \sim 1/\omega^*$ where $\omega^*$ represents a characteristic energy scale, above which the critical behavior is observed. Hence, we choose the energy gap of the lowest frequency mode $\hbar \omega_{\min} \approx \hbar \omega_c/\Lambda^M$ as the inverse of the size $1/L$. For a sufficiently large number of bath modes, e.g., $M=430$, the finite-size effect is already negligibly small.

According to the quantum-to-classical correspondence, the SBM can be mapped to a one-dimensional classical Ising chain,
\begin{equation}
\hat{H} = -\sum_{i,j}J_{ij}S_iS_j+ \hat{H}_{\rm short-range},
\label{class_hami}
\end{equation}
where $J_{ij}=J/r^{1+s}$ represents the long-range interaction with the distance $r$ between two spins, $S_i=\pm 1$ denotes the classical Ising spin, and $\hat{H}_{\rm short-range}$ is an additional generic short-range interaction irrelevant to the critical behavior. In general, it is believed that the quantum transition of SBM is in the same universality class as that of the classical model \cite{win09,chi11,guo12}. Further analysis gives the mean-field results:
\begin{equation}
\beta = 1/2, \qquad \gamma=1, \qquad  \delta=3, \qquad \nu=1/s.
\label{mean_field}
\end{equation}
In the non-mean-field regime ($0.5<s<1$), numerical results of the critical exponents can be estimated from the analytical prediction by the two-loop renormalization-group theory as well as the hyperscaling relation \cite{fis72},
\begin{eqnarray}
\label{two-order RG}
\gamma & = &  1 + \frac{4}{3}\epsilon + 4.368 \epsilon^2 + \mathcal {O}(\epsilon^3),         \nonumber \\
1/\nu &=&  \frac{1}{2} + \frac{1}{3}\epsilon - 2.628 \epsilon^2 + \mathcal {O}(\epsilon^3)      \\
\beta  & = & \gamma \frac{1-s}{2s}, \qquad \delta  =  \frac{1+s}{1-s},  \nonumber
\end{eqnarray}
with the expansion parameter $\epsilon=s-1/2$.

\section{Numerical results} \label{sec:num}
With numerical variational calculations, quantum criticality of the sub-Ohmic SBM in a high dense spectrum is investigated in this section for different values of the spectral exponent $s=0.1,0.2,\cdots, 0.9$ in the weak tunneling taking the setting of the tunneling amplitude $\Delta=0.1$ as an example. Theoretically, the number of effective bath modes $M \rightarrow \infty$ is required for the completeness of the environment. Considering the constraint available computational resources, a large value of $M = 430$ is used in main results. Similarly, the number of coherent-superposition states $N = 8$ is set. Both of them have been confirmed to be sufficient in accurately describe ground states through the convergency test \cite{zho14,zho21,zho22}. Thus, more than $8600$ variational parameters are used for each set of the model parameters in calculations. The energy bias $\varepsilon=0$ is set in the following unless otherwise noted. Statistical errors of the critical couplings and exponents are estimated by dividing the total samples into two subgroups. If the fluctuation in the curve is comparable with or larger than the statistical error, it will be taken into account.

\subsection{Estimation of critical couplings and exponents}
\begin{figure*}[htbp]
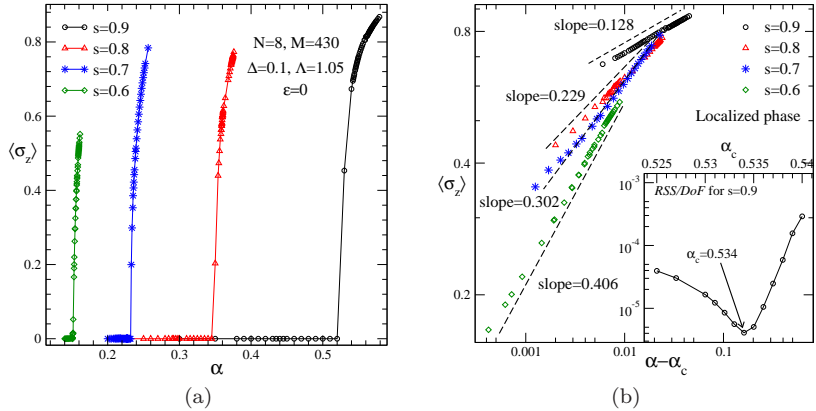

  \centering
  \subfigure[]{
   \includegraphics[scale=0.3]{magnetic.eps}
   }
   \quad
   \subfigure[]{
   \includegraphics[scale=0.18]{mag_2.eps}
   }
   \caption{ (a) The ground-state magnetization $\langle\sigma_z\rangle$ in the shallow sub-Ohmic SBM is displayed as a function of the coupling $\alpha$ for different values of the spectral exponent $s=0.6, 0.7, 0.8$, and $0.9$ in the case of $\Delta/\omega_c=0.1$ and $\varepsilon =0$. The discretization factor $\Lambda=1.05$ and numbers of coherent-superposition states and effective bath modes, $N=8$ and $M=430$, are set. (b) The log-log plot of the magnetization $\langle\sigma_z\rangle \ vs. \ (\alpha-\alpha_c)$ where dashed lines represent power-law fits. Taking $s=0.9$ as an example, inset illustrates the fitting error $RSS/DoF$ at different input value of the critical coupling $\alpha_c$, and the arrow indicates the location of the minimization. }
   \vspace{0.5\baselineskip}
   \label{f2}
\end{figure*}

In Fig.~\ref{f2}(a), the ground-state spin magnetization $\langle\sigma_z\rangle$ in the shallow sub-Ohmic SBM is displayed as a function of the coupling $\alpha$ for $s=0.6,0.7,0.8$, and $0.9$. Quantum phase transition is demonstrated from the delocalized phase ($\langle\sigma_z\rangle=0$) to the localized phase ($\langle\sigma_z\rangle \neq 0$). The transition point $\alpha_c$ is then obtained, which increases with the spectral exponent $s$. With $\alpha_c$ as input, a power-law behavior of $\langle\sigma_z\rangle$ is detected with respect to the shift $\alpha-\alpha_c$, as shown in Fig.~\ref{f2}(b). In the inset, the fitting error, as defined by the residual sum of squares over the degree of freedom ($RSS/DoF$), is carefully examined in a narrow regime of the coupling $\alpha$, taking the case $s=0.9$ as an example. It provides a measure of the average error or variability per degree of freedom. A lower value of $RSS/DoF$ indicates a better fit. By judging the location of the minimum fitting error, one determines the critical coupling $\alpha_c=0.534(1)$ which is more accurate than before. It agrees well with numerical results $0.548(2)$ and $0.555$ obtained from QMC and VMPS works \cite{win09,guo12}, respectively, showing the validity of NVM. The critical exponent $\beta$ can also be measured with Eq.~(\ref{scaling law}). For each case, the value of $\beta$ is significantly less than $1/2$, suggesting that the mean-field prediction is violated in the shallow sub-Ohmic SBM. It is in contrast to the claim obtained from the previous variational work \cite{he18}.

\begin{figure*}[htbp]
  \centering
   \includegraphics[scale=0.4]{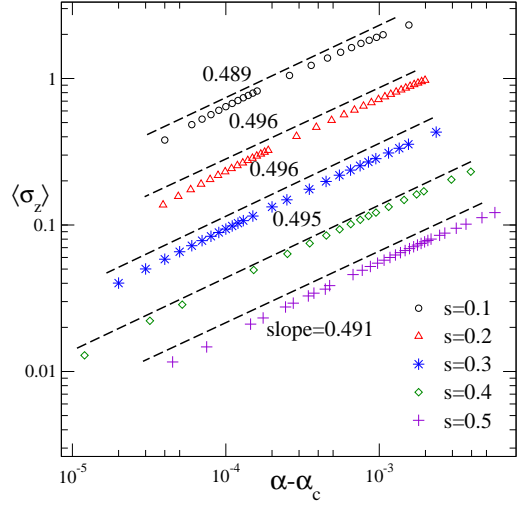}
   \caption{ In the deep sub-Ohmic regime $s\leq 0.5$, the magnetization $\langle\sigma_z\rangle$ with respect to the critical coupling deviation $\alpha - \alpha_c$ is plotted with open circles, open triangles, stars, open diamonds, and pluses, corresponding to $s=0.1,0.2,0.3,0.4$, and $0.5$, respectively. For clarity, curves are shifted downwards or upwards. Dashed lines show power-law fits. }
   \vspace{1.5\baselineskip}
   \label{f3}
\end{figure*}

For comparison, the spin magnetization $\langle\sigma_z\rangle$ in the deep sub-Ohmic SBM is plotted in Fig.~\ref{f3} for different spectral exponent $s \leq 0.5$ on a log-log scale. The power-law growing curves are almost parallel to each other, quite different from those in Fig.~\ref{f2}(b). The slopes $\beta=0.489(9),0.496(7),0.496(5),0.495(7)$, and $0.491(6)$ are then measured, corresponding to $s=0.1,0.2,0.3,0.4$, and $0.5$, respectively. All of them are in good agreement with the mean-field prediction in Eq.~(\ref{mean_field}). It confirms the robustness of the mean-field nature for quantum phase transitions in the deep sub-Ohmic regime $s<0.5$, as it is reported in literatures \cite{win09,chi11,guo12,fre13,he18}.

\begin{figure*}[htbp]
  \centering
   \includegraphics[scale=0.4]{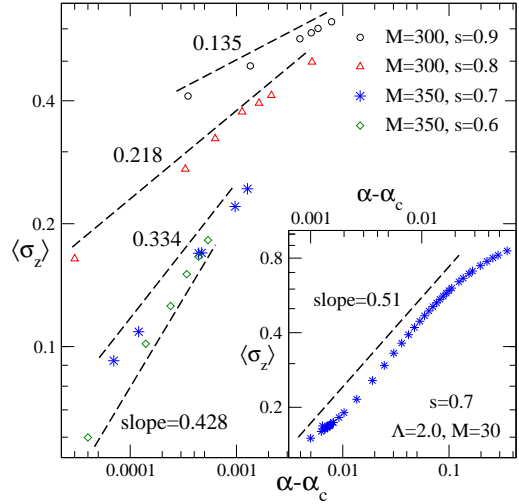}
   \caption{ Power-law behaviors of $\langle\sigma_z\rangle$ with respect to the shift $\alpha - \alpha_c$ are displayed for the sizes $M=300$ and $350$ in the cases of $s=0.9, 0.8$ and of $s=0.7,0.6$, respectively. In the inset, the discretization factor $\Lambda=2.0$ is set, much larger than the previous one $\Lambda=1.05$. Dashed lines represent power-law fits. }
   \label{f4}
\end{figure*}

In Fig.~\ref{f4}, the influence of the bath-mode number $M$ related to the size of the system $L$ is investigated, taking $M=350$ for the cases of $s=0.6$ and $0.7$, and $M=300$ for the cases of $s=0.8$ and $0.9$. The values of the critical exponent $\beta$ are measured from the slopes at different $s$, comparable with those in Fig.~\ref{f2}(b). It confirms the finite-size effect induced by the number $M$ is already immaterial in main results. In addition, calculations with a lager discretization factor $\Lambda=2$ are also performed, which was commonly employed in earlier works \cite{bul05,hur08,ber14}. Even for the case of $s=0.7$, the magnetization grows as a power law with a mean-field exponent $\beta \approx 0.5$ over two decades in the inset of Fig.~\ref{f4}, quite different from $\beta=0.128(1)$ in Fig.~\ref{f2}(b). The overestimation of $\beta$ is believed to be caused by the artificial mean-filed nature. It suggests that numerical works with a large $\Lambda$ may lead to a poor approximation of the ground state. This is the possible reason of the failure in numerical variational work \cite{he18}. Hence, the continuum limit $\Lambda \rightarrow 1$ should be taken to obtain an accurate description of quantum transitions.

\begin{figure*}[htbp]
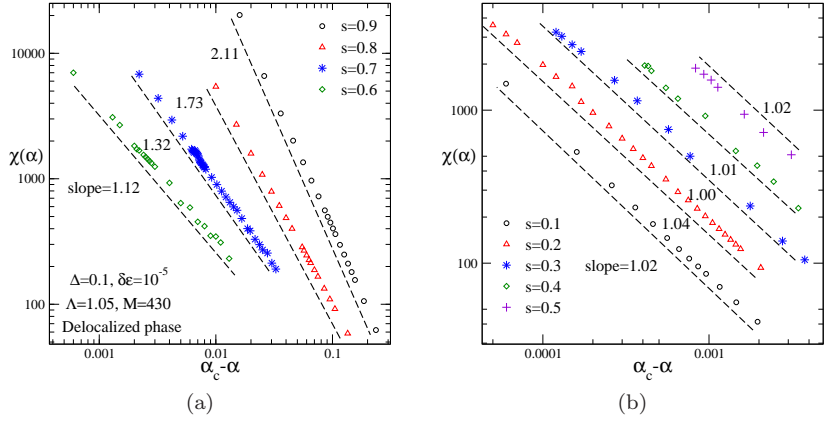

  \centering
  \subfigure[]{
   \includegraphics[scale=0.3]{chi.eps}
   }
   \quad
   \subfigure[]{
   \includegraphics[scale=0.3]{chi2.eps}
   }
   \caption{ In the delocalized phase, the local susceptibility $\chi(\alpha)$ is plotted as a function of $\alpha_c-\alpha$ for the cases in the shallow sub-Ohmic regime (a) and deep sub-Ohmic regime (b), respectively. The tiny bias $\delta\varepsilon / \omega_c = 10^{-5}$ is used to calculate the susceptibility $\chi=\partial \langle\sigma_z\rangle / \partial \varepsilon |_{\varepsilon=0}$, and all power-law decaying curves are fitted by dashed lines. }
    \vspace{2\baselineskip}
   \label{f5}
\end{figure*}

Subsequently, the extension to the biased SBM is performed for further studies. In the delocalized phase, the local susceptibility $\chi(\alpha)$ defined in Eq.~(\ref{vm_sup}) is plotted in Fig.~\ref{f5} for different $s$ under a tiny bias $\delta\varepsilon =10^{-5} \omega_c$. With the power-law scaling in Eq.~(\ref{scaling law}), the measurement of the slope yields the value of the critical exponent ranging from $\gamma=1.12(1)$ to $2.11(1)$ in the shallow sub-Ohmic regime and $\gamma \approx 1$ in the deep sub-Ohmic regime, respectively. The latter is in good agreement with the mean-field prediction in Eq.~(\ref{mean_field}), thereby again lending support to the reliability of the variational results.

\begin{figure*}[htbp]
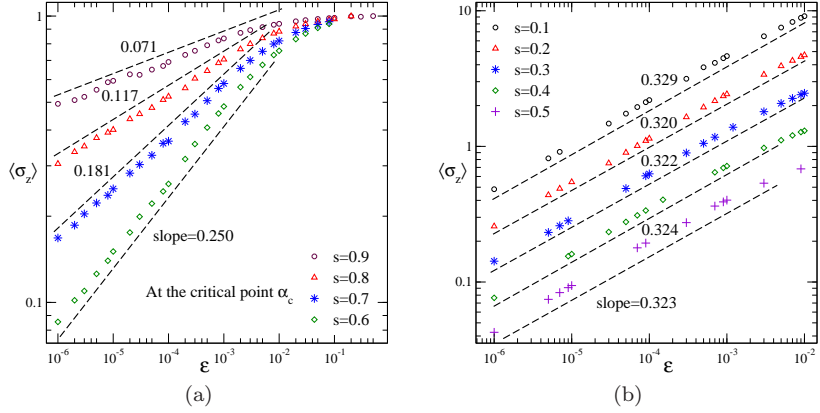

  \centering
  \subfigure[]{
   \includegraphics[scale=0.3]{field.eps}
   }
   \quad
   \subfigure[]{
   \includegraphics[scale=0.3]{field_2.eps}
   }
   \caption{ The spin magnetization $\langle\sigma_z\rangle$ at the critical point $\alpha_c$ is plotted in response to an external field $\varepsilon$ for various values of s in the shallow sub-Ohmic SBM (a) and deep sub-Ohmic SBM (b), respectively. Dashed lines show power-law fits, and the curves in (b) are shifted up for clarity. }
    \vspace{2\baselineskip}
   \label{f6}
\end{figure*}

In addition, the response of the external field (bias) at the transition point $\alpha_c$ is investigated in Fig.~\ref{f6} for different values of $s$. Non-linear increases of the magnetization $\langle\sigma_z\rangle$ are found with respect to the applied field $\varepsilon$ in both shallow and deep sub-Ohmic regimes, signifying a second-order phase transition. The critical exponent $1/\delta$ is then determined from simply fitting the slope of the curve. In the shallow sub-Ohmic regime, the value of the exponent decreases rapidly with $s$, and approaches zero in the Ohmic case. The latter indicates that the magnetization should be always unity over the whole coupling ranging of the localized phase, consistent with previous reports \cite{hur08,naz12,zho18}. In the deep sub-Ohmic one, however, the measurements of $1/\delta$ coincide with each other, showing the independence of $s$.

All the results of the transition point $\alpha_c$ for different spectral exponent $s$ are summarized in Table.~\ref{t1}, in comparison with those in the literature. Based on the extension of the Silbey-Harris polaron ansatz as well as the Ginzburg-Landau theory, the theoretical value of the critical coupling in the deep sub-Ohmic SBM has been solved analytically \cite{chi11},
\begin{equation}
\alpha_c=\frac{\sin(\pi s)e^{-s/2}}{2\pi(1-s)}\left(\frac{\Delta}{\omega_c}\right)^{1-s}.
\label{alpha_c}
\end{equation}
The numerical results are then estimated, as shown in the penultimate column of the table. However, they are significantly smaller than those obtained from general numerical methods, i.e., NRG, QMC, VMPS, and DMRG \cite{bul03,voj05,win09,guo12,won08}. It indicates that the single polaron ansatz employed in this analytical work seems not sophisticated enough to capture full bath quantum fluctuations which play a crucial role in quantum phase transitions. Moreover, the analytical solution of $\alpha_c$ is still missing in the shallow sub-Ohmic SBM. In this work, variational results of the transition point $\alpha_c$ are significantly improved by employing the NVM based on systematic coherent-state decomposition in the continuum limit. The critical couplings in both deep and shallow sub-Ohmic regimes are determined accurately, as shown in the last column. The results accord well with those obtained by QMC, VMPS, and DMRG, although having a notable deviation from the NRG findings in the shallow sub-Ohmic regime. Therefore, our numerical method, NVM, demonstrates remarkable superiority,  the same as QMC, VMPS, and DMRG. Moreover, the physical interpretation of our method is notably clearer and simpler, adding to its advantage over other approaches.

Finally, the measurements of critical exponents $\beta, \gamma$, and $1/\delta$ are summarized in Table.~\ref{t2}, in comparison with those of recent VMPS work \cite{guo12} and theoretical predictions in Eqs.~(\ref{mean_field}) and (\ref{two-order RG}). In the deep sub-Ohmic regime, the exponent remains almost unchanged with respect to the spectral exponent $s$. Compared to the VMPS ones, our NVM results are closer to the mean-field predictions $\beta=1/2, \gamma=1$, and $1/\delta=1/3$. In the shallow sub-Ohmic regime, both $\beta$ and $1/\delta$ monotonically decrease with $s$, showing the opposite tendency of $\gamma$. All of them are in good agreement with theoretical results obtained from the two-loop renormalization-group analysis, as shown in the left columns. It provides a strong confirmation on the validity of the quantum-to-classical correspondence in the entire sub-Ohmic regime. In addition, our results are also comparable with VMPS ones, again pointing to the superiority of NVM in studying quantum phase transitions.

\subsection{Other observables related to the spin and bath}

\begin{figure*}[htbp]
  \centering
   \includegraphics[scale=0.4]{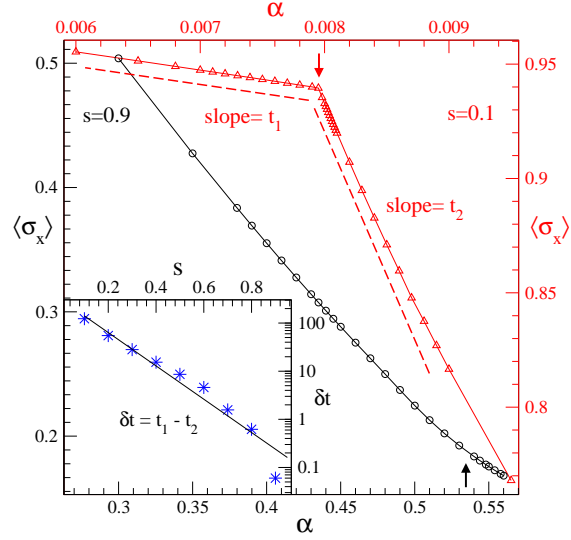}
   \caption{The spin coherences $\langle\sigma_x\rangle$ in the deep ($s=0.1$) and shallow ($s=0.9$) sub-Ohmic cases are displayed as a function of the coupling $\alpha$ in the right-top panel and left-bottom panel, respectively. As before, other parameters $\Delta/\omega_c=0.1, \varepsilon =0, \Lambda=1.05, N=8$ and $M=430$ are set. The transition points $\alpha_c=0.00795$ and $0.534$ are marked by the arrows. In the inset, the difference $\delta t=t_2-t_1$ between the slopes on two sides of the transition point is plotted against the spectral exponent $s$. Dashed lines show linear fits, and solid line in the inset represents the fit with an exponential function. }
   \label{f7}
\end{figure*}

For a better understanding of quantum phase transitions, the spin coherence $\langle \sigma_x \rangle$ is also investigated in Fig.~\ref{f7} as a function of the coupling $\alpha$ for different values of $s$. Two lines with open triangles and open circles are demonstrated, corresponding to the deep sub-Ohmic case of $s=0.1$ and shallow sub-Ohmic case of $s=0.9$, respectively. In the former, distinct values of the slopes $t_1$ and $t_2$ are measured from two sides of the transition point $\alpha_c=0.00795$, indicating a discontinuity in the slope of the first derivative of the ground-state energy $\langle \sigma_x \rangle=\partial E_g / \partial \Delta$. Therefore, the second-order phase transition is clearly evidenced in the deep sub-Ohmic SBM. In the latter, however, the spin coherence $\langle \sigma_x \rangle$ decreases smoothly, and a slight kink occurs at the transition point $\alpha_c=0.534$. Specifically, the slope difference $\delta t=t_2-t_1$ at $s=0.9$ is approximately three orders of magnitude less than that at $s=0.1$. It means that the sharp transition is significantly softened, yielding the non-mean-field critical behavior. Furthermore, an exponential decay of $\delta t(s)$ is demonstrated in the inset with the slope $7.8$. The vanishing difference $\delta t=0$ is then inferred in the limit $s \rightarrow 1$, supporting the argument that the spin coherence $\langle \sigma_x \rangle$ should be continuous at $\alpha_c$ in the presence of an Ohmic bath  \cite{ber14}.

\begin{figure*}[htbp]
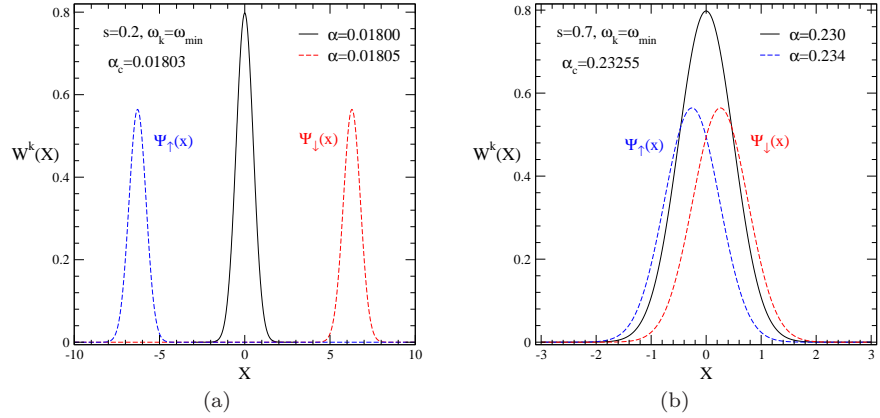

  \centering
  \subfigure[]{
   \includegraphics[scale=0.3]{win.eps}
   }
   \quad
   \subfigure[]{
   \includegraphics[scale=0.3]{win2.eps}
   }
   \caption{ Wigner distribution $W^k(X)$ which allows direct characterization of the ground-state wavefunction relevant to the equilibrium bosonic bath is shown for the delocalized phase $\alpha < \alpha_c$ and localized phase $\alpha > \alpha_c$, taking the cases of $s=0.2$ in (a) and $s=0.7$ in (b) as examples. All bath modes are traced out except the one with the lowest frequency $\omega_k=\omega_{\rm min}$, and the wigner distribution of $\Psi_\uparrow(X)$ and $\Psi_\downarrow(X)$ representing twofold-degenerate ground states are scaled by a factor $1/\sqrt{2}$ for clarity. }
   \label{f8}
\end{figure*}

Besides, ground-state properties of the bosonic bath in the vicinity of the transition point are analyzed, as depicted by the function $W^k(X)$ in Eq.~(\ref{vm_wigner}).  Two cases of $s=0.2$ and $0.7$ are considered for instance, corresponding to the deep and shallow sub-Ohmic SBM, respectively. In the Fig.~\ref{f8}(a), a significant increase in the mean position $\overline{X}=\int X W^{k}(X)dX$ is observed from the delocalized phase with $\overline{X}=0$ to the localized phase with $6.29$, as the coupling strength $\alpha$ is changed by only a paltry amount of $0.00005$. The shift in the mean position $\delta\overline{X}$ is much larger than the standard deviation $\Delta X \approx 0.5$ which represents the quantum fluctuation. It indicates the robustness and mean-field effect of the phase transition \cite{win09,chi11}, which is caused by the low-frequency nonadiabatic bath modes whose displacements diverge with the same sign as $\omega \rightarrow 0$. Additionally, two degenerate ground state $\Psi_{\uparrow}(x)$ and $\Psi_{\downarrow}(x)$ are identified, each possessing opposite displacements to the other, corresponding to the magnetization $m=\pm \langle\sigma_z\rangle$ in Fig.~\ref{f1}.

In the shallow sub-Ohmic regime, however, the criticality of the bosonic bath differs significantly from that in the deep one. The curves of Wigner distribution $W^k(X)$ mostly overlap for $\alpha=0.230$ and $0.234$ on two sides of the transition point, as shown in the subfigure (b). It highlights that the quantum fluctuation holds a crucial role even for the bath mode with the lowest frequency $\omega_{\rm min}$, resulting in a non-mean-field scaling behavior. Furthermore, the spin magnetization $\langle\sigma_z\rangle=\pm 0.299 >0$ is found in both the two-folding degenerate states, where the mean position $\overline{X}=\pm 0.257$ is less than the quantum fluctuation $\Delta X\approx0.5$. This observation implies that the order parameter is highly sensitive to the numerical errors in the ground state of the bosonic bath, providing a great challenge to obtain an accurate description of the quantum phase transition. For this reason, numerical results in earlier works \cite{bul03,alv09,win09,zha10,guo12}, including critical couplings and exponents, exhibit considerable discrepancies when the spectral exponent is close to $s=1$.

\subsection{Finite-size scaling}
\begin{figure*}[htbp]
  \centering
   \includegraphics[scale=0.3]{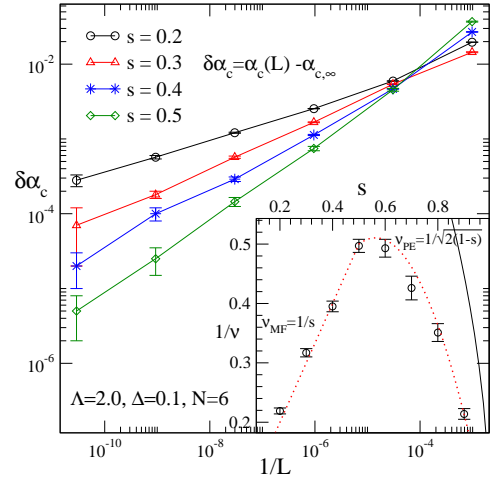}
   \caption{The shift of the transition point $\delta\alpha_c=\alpha_c(L)-\alpha_{c,\infty}$ is displayed against the inverse of the size $1/L$ for different values of the spectral exponent $s$ on a log-log scale. The critical exponent $1/\nu$ extracted from the finite-size scaling behavior $\delta\alpha_c \sim L^{-1/\nu}$ is shown in the inset as a function of $s$. For comparison, the theoretical prediction of $1/\nu$ is also given with the dotted line which is obtained from the mean-field ($s \leq 0.5$) and two-loop renormalization-group ($s>0.5$) methods. Moreover, solid line in the inset represents the perturbative prediction $\nu=1/\sqrt{2(1-s)}$ when $s$ close to $1$. }
   \label{f9}
\end{figure*}

Subsequently, the correlation-length exponent $\nu$ describing the divergence of the correlation length $\xi$ is investigated by the finite-size scaling. Specifically, the dependence of the critical coupling $\alpha_c$ on the bath size $L=1/\omega_{\rm min}$ is investigated, and the results are shown in Fig.~\ref{f9} where the discretization factor $\Lambda=2.0$ is set for convenience. Power-law dependence of the transition shift, $\delta \alpha_c \sim (1/L)^{1/\nu}$, are found on the bath size not only in the deep sub-Ohmic regime for $s=0.2,0.3,0.4$ and $0.5$, but also in the shallow one for $s=0.6,0.7,0.8$ and $0.9$ (not shown). A direct measurement of the slope gives the value of the exponent $1/\nu$ which varies with the spectral exponent $s$.

As shown in the inset, the exponent $1/\nu$ exhibits a linear increase when $s \leq 0.5$, in excellent agreement with the mean-field prediction $1/\nu_{_{MF}}=s$. For $s>0.5$, however, it decreases with $s$ and vanishes in the Ohmic case $s=1$, showing the divergence of the correlation-length exponent $\nu$. Moreover, our results closely follow the predictions of the two-loop renormalization-group method within the error bars, but are much leas than those from the perturbative calculations $1/\nu_{_{PE}}=\sqrt{2(1-s)}$. It indicates that the formula in the latter holds only when $s>0.9$, which is much closer to the Ohmic case.

\section{Conclusion}\label{sec:con}

With large-scale numerical variational calculations, comprehensive studies have been performed on quantum phase transitions in dissipative quantum systems, taking the sub-Ohmic spin-boson model with a highly dense spectrum as an example. In these variational calculations, a generalized trial wave function composed of coherent-state expansions has been employed to effectively capture quantum entanglements and quantum fluctuations in the bath which play a crucial role in determining the quantum phase transition. Thus, systematical investigations into the quantum criticality of both the spin and bosonic bath have been carried out, leading to the accurate determination of the transition point $\alpha_c$ as well as the critical exponents $\beta, \nu, \gamma$, and $1/\delta$  for various spectral exponents $s=0.1,0.2,\cdots, 0.9$ with a fixed tunneling constant of $\Delta=0.1$.

In contrast to previous variational results based on the single polaron ansatz, our results of the critical couplings demonstrate excellent agreement with those obtained by the QMC, VMPS, and DMRG methods. Although a minor discrepancy is observed when comparing with the NRG results in the shallow sub-Ohmic regime. It thereby lends support to the validity of the variational calculations in this work. Additionally, the quantum-to-classical correspondence has been fully confirmed over the entire sub-Ohmic range, for the consistency between the numerical results of the critical exponents in the SBM and  theoretical predictions of one-dimensional classical Ising spin chain with long-range interactions which are derived from the two-loop renormalization-group analysis and the hyperscaling relation. Finally, two distinct influences of quantum fluctuations have been uncovered on the low-frequency bath modes in the deep and shallow sub-Ohmic regimes, respectively, which may be responsible for the corresponding manifestation of mean-field and non-mean-field critical behaviors.

As the supplementary material, we supply the FORTRAN codes for the numerical variational method, entitled ``init-bath.f90'' and ``pre-bath.f90'' corresponding to Steps 1-3 and Steps 4-5, respectively, to help further development in this field.

{\bf Acknowledgements:} This work was supported by National Natural Science Foundation of China under Grant Nos. $11875120$.

\bibliographystyle{apsrev4-1}
\bibliography{sbm}

\newpage

\begin{table*}[htbp]
\small
\centering
\tabcolsep=0.1cm
\renewcommand\arraystretch{1.5}
\begin{tabular}[t]{r| c  c c c  c c}
\hline
\hline
$s$           &     NRG \cite{bul03,voj05}          &   QMC \cite{win09}            &   VMPS \cite{guo12}            &      DMRG \cite{won08}     &   Single polaron \cite{chi11} & This work     \\
\hline
0.1           &   0.008(1)       &   0.0076(3)      &   $\backslash$    &     0.0074(2)      &  0.0065          &   0.00795(1)  \\
0.2           &   0.018(1)       &   0.0175(2)      &   0.0175367       &     0.0162(2)      &  0.0168          &   0.01806(3)  \\
0.3           &   0.035(2)       &   0.0350(5)      &   0.0346142       &     0.0332(5)      &  0.0316          &   0.03470(6)  \\
0.4           &   0.064(2)       &   0.0604(8)      &   0.0605550       &     0.058(1)       &  0.0519          &   0.0604(2)   \\
0.5           &   0.106(2)       &   0.098(1)       &   0.0990626       &     0.099(1)       &  0.0784          &   0.0977(3)   \\
0.6           &   0.170(2)       &   0.157(2)       &   0.1554073       &     0.155(1)       &  $\backslash$    &   0.1528(3)   \\
0.7           &   0.261(2)       &   0.241(2)       &   $\backslash$    &     0.238(2)       &  $\backslash$    &   0.2345(8)   \\
0.8           &   0.392(3)       &   0.360(2)       &   $\backslash$    &     0.359(2)       &  $\backslash$    &   0.354(1)   \\
0.9           &   0.612(3)       &   0.548(2)       &   0.5555478       &     0.556(2)       &  $\backslash$    &   0.534(1)    \\
\hline \hline
\end{tabular}
\caption{ The critical coupling $\alpha_c$ in the sub-Ohmic SBM is shown for different values of the spectral exponents $s$. The last two columns correspond to those obtained by NVM on the basis of single polaron ansatz ($N=1$) and multiple poalron ansatz ($N=8$), respectively. With the plotdigitizing technique, the NRG, QMC, and DMRG results are extracted from the figures in literatures, and the corresponding error bar is estimated as the maximum inaccuracy of the readout.  }
\label{t1}
\end{table*}

\begin{table*}[htbp]
\small
\centering
\tabcolsep=0.1cm
\renewcommand\arraystretch{1.5}
\begin{tabular}[t]{r | c c c | c c c | c c c}
\hline
\hline
\multirow{2}{*}{$s$} &  \multicolumn{3}{c|}{ Theoretical values \cite{fis72}} &   \multicolumn{3}{c|}{ Our work }  &  \multicolumn{3}{c}{ VMPS results \cite{guo12}}  \\
\cline{2-10}  & $\beta$ &  $\gamma$ & $1/\delta$ & $\beta$ &  $\gamma$ & $1/\delta$ & $\beta$ &  $\gamma$ & $1/\delta$  \\
\hline
0.1 &   1/2 &    1   &    1/3    &  0.489(9)   &   1.02(1) &   0.329(2) &  $\backslash$ &   $\backslash$   &   $\backslash$   \\
0.2 &   1/2 &    1   &    1/3    &  0.496(7)   &   1.04(2) &   0.320(2) &  0.47(7) &  $\backslash$  & 0.39(8) \\
0.3 &   1/2 &    1   &    1/3    &  0.496(5)   &   1.00(2) &   0.322(3) &  0.50(3) &  $\backslash$  & 0.34(2)      \\
0.4 &   1/2 &    1   &    1/3    &  0.495(7)   &   1.01(2) &   0.324(7) &  0.50(3) &  $\backslash$  & 0.33(1) \\
0.5 &   1/2 &    1   &    1/3    &  0.491(6)   &   1.02(2) &   0.323(7) &  0.46(3) &  $\backslash$  & 0.30(1) \\
0.6 & 0.392 & 1.18   &    0.250  &  0.406(4)   &   1.12(1) &   0.250(3) &  0.38(1) &  $\backslash$  & 0.244(6) \\
0.7 & 0.309 & 1.44   &    0.176  &  0.302(4)   &   1.32(1) &   0.181(2) & 0.292(3) &  $\backslash$  & 0.171(3)  \\
0.8 & 0.224 & 1.79   &    0.111  &  0.229(3)   &   1.73(1) &   0.117(2) & 0.211(2) &  $\backslash$  & 0.109(2)   \\
0.9 & 0.124 & 2.23   &    0.053  &  0.128(1)   &   2.11(1) &   0.071(4) & 0.132(1) &  $\backslash$  & 0.0523(8) \\
\hline \hline
\end{tabular}
\caption{ The critical exponents $\beta, \gamma$, and $\delta$ for different $s$ are measured from extensive calculations with NVM, compared with those in the VMPS work and theoretical analysis. The latter are obtained from the analytical work based on the mean-field ($s<0.5$) and two-loop renormalization-group theories ($s>0.5$) as well as the hyper-scaling law. }
\label{t2}
\end{table*}

\end{document}